\documentclass[12pt]{article}
\usepackage{graphicx}
\usepackage{epsfig}
\usepackage{latexsym}
\usepackage{latexsym}
\usepackage{amssymb}
\usepackage{amsmath}
\usepackage{cite}
\newcommand{\reef}[1]{(\ref{#1})}

\newcommand{\ud} {\mathrm{d}}

\DeclareSymbolFont{AMSb}{U}{msb}{m}{n}
\DeclareMathSymbol{\IN}{\mathbin}{AMSb}{"4E}
\DeclareMathSymbol{\IZ}{\mathbin}{AMSb}{"5A}
\DeclareMathSymbol{\IR}{\mathbin}{AMSb}{"52}
\DeclareMathSymbol{\Q}{\mathbin}{AMSb}{"51}
\DeclareMathSymbol{\II}{\mathbin}{AMSb}{"49}
\DeclareMathSymbol{\IC}{\mathbin}{AMSb}{"43}
\DeclareMathSymbol{\IP}{\mathbin}{AMSb}{"50}
\DeclareMathSymbol{\IH}{\mathbin}{AMSb}{"48}
\DeclareMathSymbol\IA{\mathalpha}{AMSb}{"41}
\DeclareMathSymbol\IS{\mathalpha}{AMSb}{"53}

\def\Q{{\cal Q}}

\begin{document}

\begin{flushright}
DCPT-05/57
\end{flushright}
\begin{center} {\Large \bf Exactly Solvable}

\bigskip

{\Large\bf Supercritical String Theories?}

\end{center}

\bigskip \bigskip \bigskip

\centerline{\bf James Carlisle}

\bigskip
\bigskip

  \centerline{\it Centre for Particle Theory}
\centerline{\it Department of Mathematical
    Sciences}
\centerline{\it University of Durham}
\centerline{\it Durham DH1 3LE, England, U.K.}

\bigskip
\bigskip

\centerline{\small \tt j.e.carlisle@durham.ac.uk}
\centerline{\small \tt jcarlisle@supanet.com}

\bigskip
\bigskip


\begin{abstract}

By analytically continuing the string equations of the subcritical Type 0A $(2, 4|m|)$ minimal string theories, we reveal a whole new family of differential and integro-differential equations associated with the naively supercritical $(2, -4|m|)$ theories. We uncover an elegant structure, associated with the negative KdV hierarchy, that in principle yields the exact partition functions of the models for all values of the string coupling. Furthermore, the physics associated with the new equations displays many of the salient features associated with the original subcritical models, plus other new phenomena that are not present in those cases. One such phenomenon may have an interpretation as a tachyon condensation process by which the theories can change their dimensionalities. 


\end{abstract}
\newpage \baselineskip=18pt \setcounter{footnote}{0}


\section{Introduction}

Whilst much progress has been made in understanding string theory in its critical dimension, much less is understood about string theories with dimensionalities lying outside this critical value. In these \emph{noncritical} string theories the Liouville mode, $\varphi$, of the worldsheet metric remains in the model as an extra spacetime coordinate \cite{Ginsparg:1993is}. One solution to the background field equations is flat spacetime with a linear dilaton, $\Phi \propto \varphi$. These particular theories can therefore be described by Liouville gravity coupled to a worldsheet conformal field theory. They correspond to bosonic string theories living in $c+1$ dimensions, where $c$ is the central charge of the conformal field theory. Theories with $c < 25$ are known as \emph{subcritical}\footnote{It is actually very difficult to understand the $1<c<25$ regime. So in the context of this paper `subcritical' will usually refer to the $c\leq1$ models.}; whereas those with $c>25$ are known as \emph{supercritical}. We will consider the former models first. The linear dilaton means that the effective string coupling is proportional to $e^{Q \varphi /2}$, where:
\begin{eqnarray} \label{eq:Q}
Q = \sqrt{\frac{25-c}{3}}  .
\end{eqnarray}
So along the spacelike Liouville direction one sees that $\varphi \rightarrow -\infty$ corresponds to weak coupling, whereas $\varphi \rightarrow \infty$ corresponds to strong coupling. The theories permit two types of D--brane \cite{Zamolodchikov:2001ah, Fateev:2000ik, Teschner:2000md}: the FZZT branes, which stretch along the Liouville dimension from $\varphi=-\infty$ and then dissolve at some typical distance set by the boundary cosmological constant on the brane, stretching all the way to $\varphi=\infty$ only if this constant is zero; and the ZZ branes, which are fully localised in the strong coupling regime at $\varphi=\infty$. Note that in the case of $c=25$ we have $Q=0$, and so the string coupling is spatially homogeneous. In this case the Liouville direction can be Wick rotated and we recover the standard case of twenty--six dimensional critical string theory. 

One noncritical string theory that has been much studied is the two-dimensional ($c \!=\! 1$) case \cite{Klebanov:1991qa, Ginsparg:1993is}. In this model the physical state conditions remove all the string excitations except the tachyon, which is actually massless in two dimensions. It turns out that contributions to the partition function are heavily suppressed in the strong coupling $\varphi \rightarrow \infty$ region, and so strings are prevented from exploring it. Therefore the physics of the model consists of tachyons propagating from $\varphi = -\infty$ towards $\varphi = \infty$ until they reach some typical distance, known as the \emph{Liouville wall}, at which point they begin to interact strongly and scatter back towards $\varphi = -\infty$. An appropriate S-matrix can be defined for such a process.

Other important noncritical string theories arise from the coupling of the minimal conformal models to Liouville gravity \cite{Staudacher:1989fy, Brezin:1989db}. These are the \emph{minimal string theories} living in less than two target spacetime dimensions. In many of these minimal string theories it is possible to write down differential equations that, in principle, allow exact solutions to be computed for the partition function at all values of the string coupling. The models are labelled by two positive coprime integers $(p,q)$ with central charges given by:
\begin{eqnarray}
c = 1 - \frac{6(p-q)^2}{pq}  .
\end{eqnarray}
So an example would be the $(2,3)$ model corresponding to $c=0$ pure worldsheet gravity. 

In addition to the above bosonic theories, one can also define supersymmetric noncritical string theories. These correspond to super-Liouville gravity coupled to superconformal field theories \cite{Distler:1989nt}. Just like in the bosonic case we can define theories related to the $(p,q)$ superconformal minimal models, where this time $p$ and $q$ must be either both odd and coprime; or both even with $p/2$ and $q/2$ coprime. In the latter case one must also have $(p-q)/2$ odd \cite{Seiberg:2003nm}. The central charge, $\hat{c}$, and the analogue of $Q$ in \reef{eq:Q}, of these superminimal models are given by \cite{Distler:1989nt}:
\begin{eqnarray} \label{eq:Supercharge}
\hat{c} = 1 - \frac{2(p-q)^2}{pq}  , \qquad Q = \sqrt{\frac{9-\hat{c}}{2}}  .
\end{eqnarray}
We will see below that the $(2,4m)$ ($m=1,2,\dots$) series of $\hat{c}<1$ Type 0A superstring theories is intimately related to the KdV hierarchy of integrable equations\footnote{We will also see that the same integrable hierarchy underpins the bosonic $(2,2m-1)$ theories.}.
\\

In comparison to the subcritical models, relatively little is known about supercritical string theories living in greater than the critical number of dimensions ($c>25, \, \hat{c}>9$). Some progress has been made however, such as in \cite{Myers:1987fv, Antoniadis:1990uu, Polchinski:1989fn, Cooper:1991vg, DaCunha:2003fm, Martinec:2003ka}, but it would be very useful to learn more. Once again one finds that a solution to the background field equations is the linear dilaton. However, the parameter $Q$ in \reef{eq:Q} and \reef{eq:Supercharge} is now imaginary, which means that the Liouville direction must be timelike in these theories. This will lead to a theory that is either strongly coupled in the far past or strongly coupled in the far future. Consequently the theory is unstable and it is not clear how to define an S-matrix. In an attempt to overcome these instabilities, other solutions to the background field equations have been suggested \cite{Silverstein:2001xn, Maloney:2002rr}, but there is still much to be understood. 
\\

\begin{figure}[ht]
\begin{center}
\includegraphics[scale=0.55]{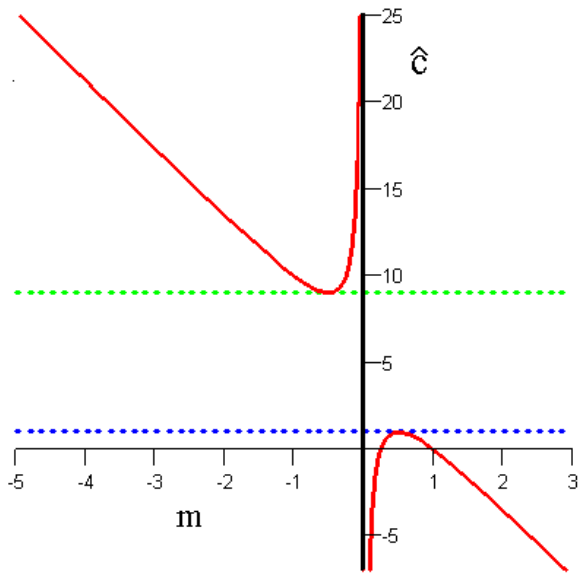} 
\end{center}
\caption{\small The central charge, $\hat{c}$, of the $(2,4m)$ Type 0A string theory is plotted against $m$ (red). Also shown are the lines $\hat{c}=1$ (dotted blue) and $\hat{c}=9$ (dotted green). By analytically continuing to negative values of $m$ it may be possible to escape from the subcritical $(\hat{c} \leq 1)$ region and break through to the supercritical ($\hat{c} > 9$) region.} 
\label{fig:cplot}
\end{figure}

The outline of this paper is as follows. In Section \ref{sec:Sub} we will review the many properties of the $(2,4|m|)$ series of less than two dimensional noncritical Type 0A superstring theories. This will serve as a useful comparison to what will follow. Then, in Section \ref{eq:Super}, we will attempt to extend the formalism in a way that will allow us to access a whole range of models that will naively correspond to the $(2,-4|m|)$ series. This will be a formal analytical continuation. Using \reef{eq:Supercharge} we see that the central charge of the $(2, 4m)$ models is given by:
\begin{eqnarray} \label{eq:Supercharge2}
\hat{c} = 5 - 4m - \frac{1}{m}  .
\end{eqnarray}
If we can make sense of these models for negative values of $m$ via our analytical continuation, then we see that they are expected to describe supercritical string theories. For instance, $m=-1$ corresponds to $\hat{c}=10$. This is illustrated in Figure \ref{fig:cplot}. We will see below that the string equations uncovered via this approach do yield apparently physically sensible results. Moreover, the models have some intriguing properties that may indeed relate them to supercritical theories in some way. Most interesting among these is that the $|m|$-th model naturally contains all the solutions to the $(|m|-1)$-th model. This could be a sign that a supercritical string theory of a given dimensionality is permitted to tachyon condense down to a theory of lower dimensionality as is suggested in \cite{Hellerman:2004zm, Hellerman:2004qa}. However, we should state from the outset that the main thrust of this work is the mathematical structure, which is fascinating in its own right. Any physical interpretation of these models will be conjectural by its very nature, and so we will save such speculation until Section \ref{sec:Discuss}.

\section{Subcritical String Theory} \label{sec:Sub}

In the early 1990s matrix model technology was used to formulate \emph{string equations} that in principle could be solved to yield \emph{exact} partition functions of the bosonic $(2, 2m-1)$ ($m=1,2, \dots$) minimal string theories \cite{Gross:1989vs, Douglas:1989ve, Brezin:1990rb, Dalley:1991qg}. By discretising the string worldsheet it proved possible to relate these models to the multicritical points of a Hermitian $N \times N$ matrix model \cite{Gross:1989vs, Douglas:1989ve, Brezin:1990rb}. We recover the continuum string theory by taking the \emph{double scaling limit} of sending $N \rightarrow \infty$ and simultaneously fine tuning the potential. When we do this we find that we are left with a differential equation for a function $u(z)$, which is related to the free energy (connected string partition function), $F$, of the string theory by:
\begin{eqnarray}
F = \nu^2 \frac{\partial^2 u}{\partial z^2}  .
\end{eqnarray}
The parameters $\nu$ and $z$ here are remnants of the double scaling: the former being related to $1/N$; the latter to the matrix model potential. In string theory, $\nu$ has an interpretation as the string coupling and $z$ has an interpretation as the coefficient of the operator of lowest dimension in the worldsheet conformal field theory (in the $(m, m+1)$ unitary theories it is the bulk cosmological constant). However, in general it will be possible to combine $\nu$ and $z$ into a single dimensionless parameter that will act as the renormalised string coupling. We will see this below. The string equation for the $m$--th multicritical point of the Hermitian matrix model can be written in the following compact form\footnote{In what follows we will often use the following shorthand notation: a prime will denote either $\partial \equiv \partial/\partial x$ or $\ud \equiv \nu \partial/\partial z$ depending on the context (which should always be self-evident). This will allow us to suppress $\nu$ in many of our equations. To this end we also define $\tilde{z} \equiv z/\nu$.}:
\begin{eqnarray} \label{eq:OldString}
\mathcal{R}[u] = 0  ,
\end{eqnarray}
where $\mathcal{R}[u]$ is some differential polynomial in $u(z)$ depending on the model we wish to study. For the $m$--th multicritical point we find that $\mathcal{R}$ is given by:
\begin{eqnarray} \label{eq:Bugg}
\mathcal{R} = \sum_{n=1}^m \left(n+\frac12 \right) \tilde{t}_n R_n - z ,
\end{eqnarray}
where the $\tilde{t}_n$ are constants that serve to perturb the $m$--th model with contributions from the lower models. They can be thought of as the coefficients of other operators in the worldsheet conformal field theory \cite{Dalley:1991qg}. The $R_n$ in \reef{eq:Bugg} are generated from $R_0=1/2$ via the following recursion relation:
\begin{eqnarray} \label{eq:StringRecursion}
R^{\prime}_{n+1} = (\ud^3 - 4 u \ud - 2 u') R_n  .
\end{eqnarray}
What is remarkable about this result is that this is the same mathematical structure as underlies the KdV hierarchy of integrable systems \cite{Das, Dickey}:
\begin{eqnarray} 
\alpha_m \frac{\partial u}{\partial t_m} = R_{m+1}^\prime[u]  ,
\end{eqnarray}
with $\alpha_m$ a constant. The first few $R_m$ are given by:
\begin{eqnarray} \label{eq:DiffPollies}
R_0 = \frac{1}{2}  , \quad R_1 = - u  , \quad R_2 =  3 u^2 - u''  , \nonumber \\ R_3 = 20 u' u'' + 10 u u''' - u^{(5)} - 30 u^2 u'  .
\end{eqnarray}
Note that in the case of $m=1$ we recover the original KdV equation itself:
\begin{eqnarray}
\frac{\partial u}{\partial t} = 6 u \frac{\partial u}{\partial x} -  \frac{\partial^3 u}{\partial x^3} ,
\end{eqnarray}
where we will always use the variables $x$ and $t$ in the KdV context. 
\\

Unfortunately, the above bosonic string equation \reef{eq:OldString} has non-perturbative problems. In an attempt to overcome these, the complex matrix model was studied \cite{Morris:1990cq, Morris:1990bw}. In particular, the double scaling limit of the rectangular $N \times (N+\Gamma)$ complex matrix model \cite{Myers:1992dq, Lafrance:1993wy} yields the following string equation, first written down in \cite{Dalley:1992br}:
\begin{equation} \label{eq:DJM}
u{\cal R}^2-\frac{1}{2}{\cal R}{\cal R}^{''}+\frac{1}{4}({\cal R}^{'})^2 =\nu^2\Gamma^2 ,
\end{equation}
where the parameter $\Gamma$ turns out to represent the number of background ZZ branes in the theory\footnote{More precisely it represents the number of units of ZZ brane charge in the background, and hence it can be positive or negative. When we take the squareroot of $\Gamma^2$ we will define the positive sign choice as `positive $\Gamma$' and the negative sign choice as `negative $\Gamma$' for ease of reference.}. In 2003 the string equation of this complex matrix model was reinterpreted in terms of the $(2, 4m)$ minimal superconformal theory coupled to super-Liouville theory \cite{Klebanov:2003wg}. This is Type 0A superstring theory (supersymmetric on the worldsheet, bosonic in spacetime). In the same paper the authors identified string equations arising from certain unitary matrix models as giving the partition functions of the $(2,4m)$ Type 0B superstring theories. The simplest theory is the $(2,4)$ model corresponding to pure supergravity ($\hat{c}=0$). In this case $\mathcal{R}[u] = u-z$. The resulting string equation \reef{eq:DJM} is non-linear and has the following large--$z$/weak coupling asymptotics:
\begin{eqnarray} \label{eq:UExpandPos}
u&=& z+\frac{\nu\Gamma}{z^{1/2}}-\frac{\nu^2\Gamma^2}{2z^2}+\frac{5}{32}\frac{\nu^3 \Gamma(4\Gamma^2+1)}{z^{7/2}}+\cdots \qquad \mbox{as}\quad z\to+\infty , \\
u&=& \frac{\nu^2(4\Gamma^2-1)}{4z^2}+\frac{\nu^4}{8}\frac{(4\Gamma^2-1)(4\Gamma^2-9)}{z^5}+\cdots \qquad \mbox{as}\quad z\to-\infty , \label{eq:UExpandNeg}
\end{eqnarray}
which correspond to the following free energies:
\begin{eqnarray}
F&=&\frac{1}{6}g_s^{-2}+ \frac{4}{3}\Gamma g_s^{-1}+\frac{1}{2}\Gamma^2 g_s^0\ln \, z+\frac{1}{24}\Gamma(4\Gamma^2+1)g_s^1+\cdots\qquad \mbox{as}\quad z\to+\infty ,\label{eq:forwardfree}\\
F&=&-\left(\Gamma^2-\frac{1}{4}\right)g_s^0\ln \, z+\frac{1}{96}(4\Gamma^2-1)(4\Gamma^2-9)g_s^2+\cdots\qquad \mbox{as}\quad z\to-\infty ,
\end{eqnarray}
where we have (as promised) defined a dimensionless string coupling $g_s = \nu/z^{3/2}$. A worldsheet with $b$ boundaries and $h$ handles (Euler number $\chi=2-2h-b$) is expected to be weighted by a factor of $\Gamma^b g_s^{2h+b-2}$ in the free energy. So the first term in the positive $z$ free energy expansion\footnote{In general, the $z\to+\infty$ expansion of the $(2,4m)$ model has $z^{1/m}$ leading order behaviour.} corresponds to the sphere, the second to the disc (i.e.\! an open string having one boundary ending on the ZZ branes), etc. At large negative $z$ the free energy contains only surfaces with an even Euler number. The interpretation of this regime is that $\Gamma$ represents the insertions of half-units of RR flux. This leads to the idea of a geometric transition whereby in one weak coupling regime we have open strings, closed strings and D--branes and in another we have only closed strings and fluxes. So remarkably, the relatively simple string equation \reef{eq:DJM} can in principle calculate the partition function to all orders in perturbation theory and yet still demonstrate some highly non-trivial phenomena. 

\begin{figure}[ht]
\begin{center}
\includegraphics[scale=0.55]{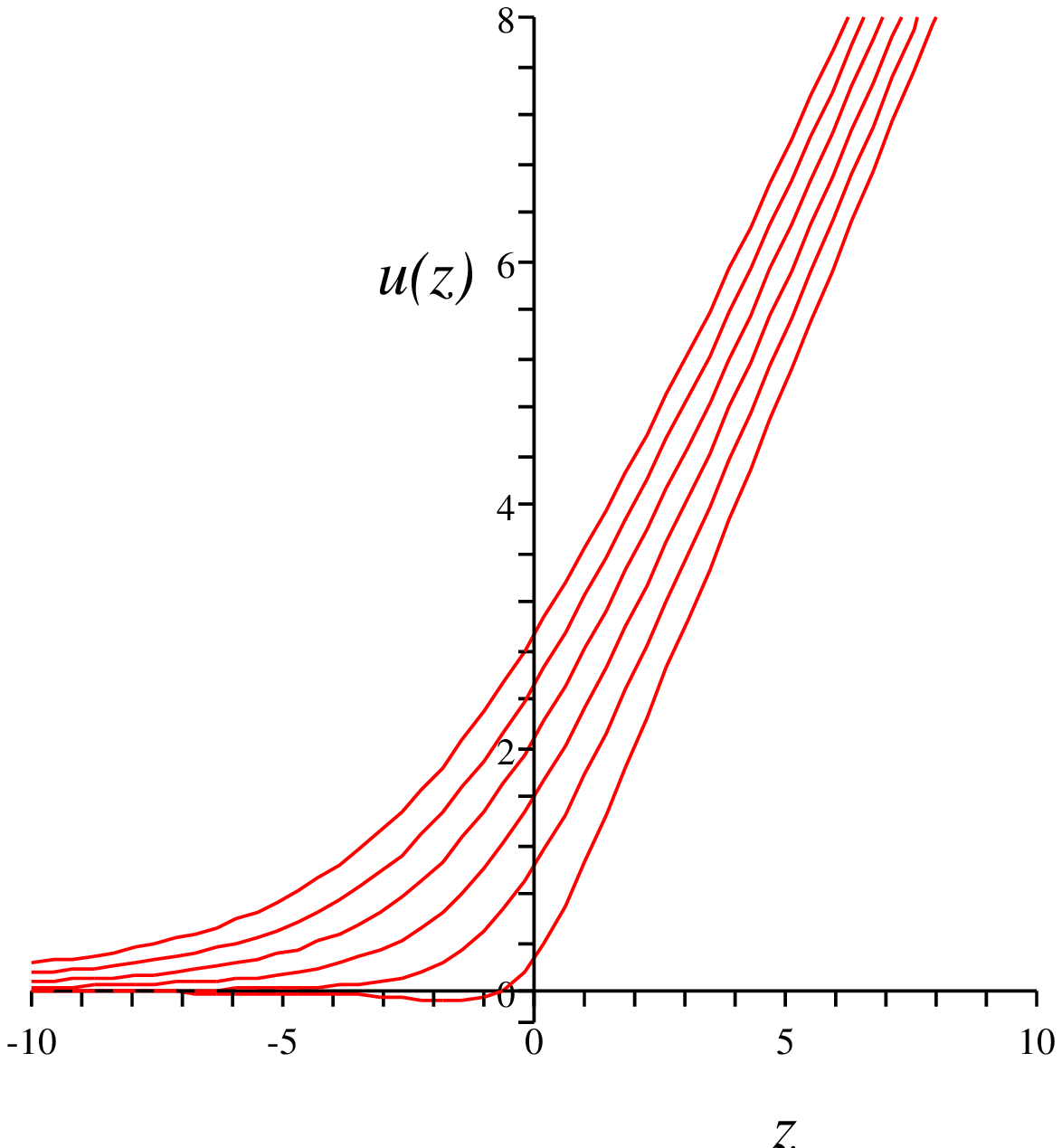} \includegraphics[scale=0.55]{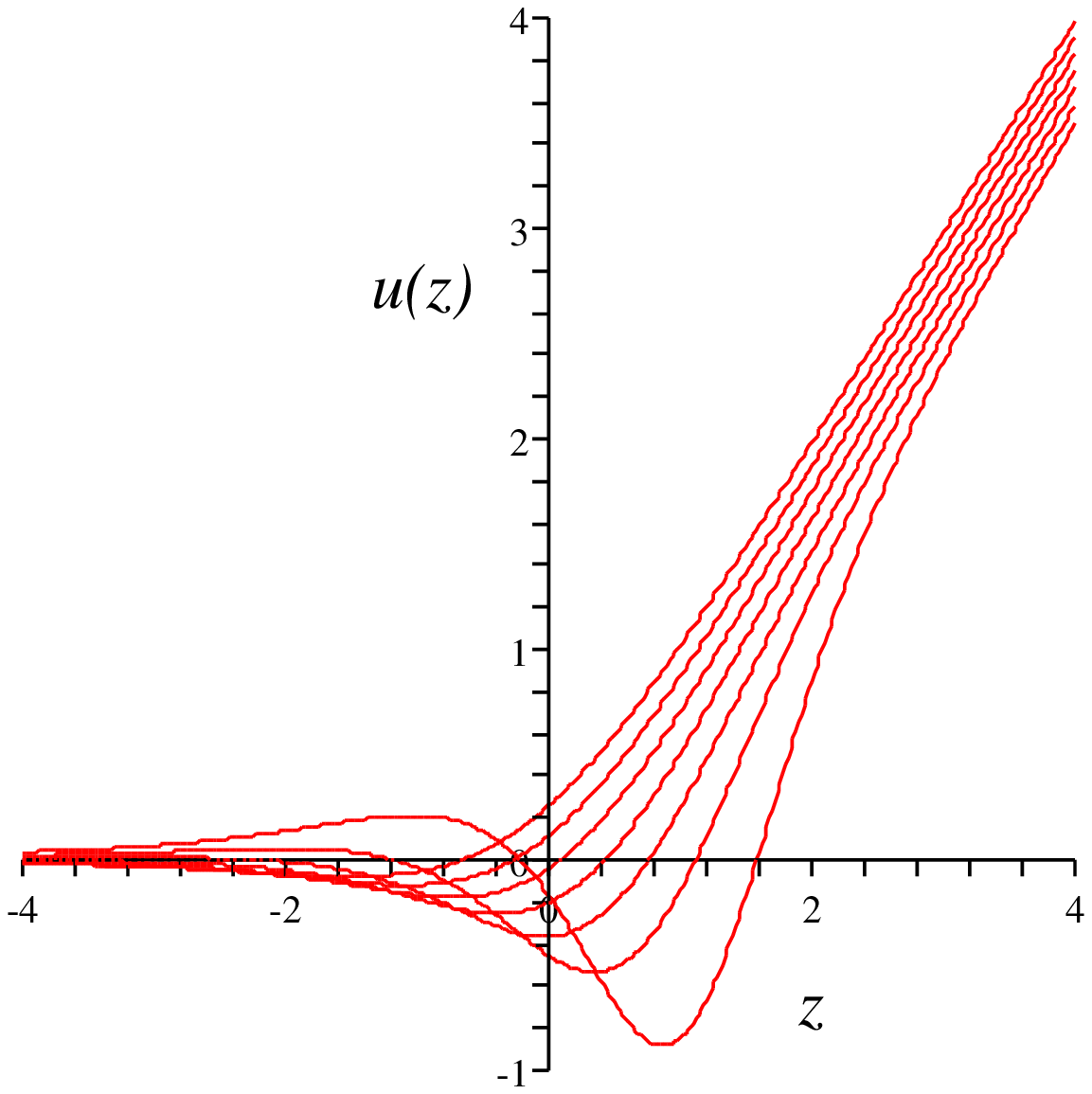} 
\end{center}
\caption{\small Numerical solutions to equation~\reef{eq:DJM} 
for $u(z)$ in the case of pure supergravity ($\mathcal{R} = u-z$): {\it (l)} cases of positive integer $\Gamma$;
 and {\it (r)} some  cases of $-1<\Gamma<0$.} 
\label{fig:gammaplots}
\end{figure}

We can do better than that though, because \reef{eq:DJM} can be solved numerically as we demonstrate for various values of $\Gamma$ in Figure~\ref{fig:gammaplots}. So technically we can compute the partition function for all values of the string coupling. Since $\Gamma$ controls the number of ZZ branes we expect it to be quantised. This matter is discussed at length in \cite{Seiberg:2004ei, Carlisle:2005mk,Carlisle:2005wa}.

The KdV hierarchy can be formulated in terms of pseudodifferential operators \cite{Douglas:1989ve, Gelfand:1976A, Douglas:1989dd}. The $m$--th member of the hierarchy is equivalent to the following wave equations in terms of the Lax pair of operators $Q$ and $P_m$:
\begin{eqnarray} \label{eq:WF1}
Q \psi \equiv (\partial^2 - u) \psi &=& \lambda \psi  , \\ \label{eq:WF2}
P_m \psi \equiv \left[ Q^{m+1/2} \right]_{+} \psi &=& \frac{\partial \psi}{\partial t_m}  ,
\end{eqnarray}
where we have formally raised the operator $Q$ to a half-integer power, which can be written as a series in powers of $\partial$ and $\partial^{-1}$ (where the latter can be formally defined as well). The `$+$' subscript refers to dropping all the terms with negative powers of $\partial$. To illustrate, $P_1$ is given by:
\begin{eqnarray}
P_1 = \partial^3 - \frac{3}{2} u \partial - \frac{3}{4} u'  .
\end{eqnarray}
We can recover the KdV equations by eliminating $\psi$ between \reef{eq:WF1} and \reef{eq:WF2}. By writing $v = -\psi'/\psi$ in \reef{eq:WF1} we obtain the so-called \emph{Miura} map, relating solutions of the KdV hierarchy to solutions of the mKdV hierarchy:
\begin{eqnarray} \label{eq:Miura}
u = v^2 - v' - \lambda  .
\end{eqnarray}
In the string theory, this relation takes on a different role. $v(z)$ turns out to be the differential of the free energy associated with a probe FZZT brane in the string theory \cite{Seiberg:2004ei, Maldacena:2004sn}. The parameter $\lambda$ plays the role of the boundary cosmological constant on the brane. We will set it to zero for our purposes, though the non-zero case is very interesting \cite{Carlisle:2005wa}. Let us illustrate the role of $v(z)$ by calculating its large--$z$ expansion using \reef{eq:UExpandPos} and \reef{eq:UExpandNeg}. We find that there are two solutions, which we shall call $v(z)$ and $\bar{v}(z)$ respectively:
\begin{eqnarray}
v &=&  z^{1/2}+\frac{1}{4}\frac{\nu (1 + 2 \Gamma)}{z}-\frac{1}{32}\frac{\nu^2}{z^{5/2}}(12\Gamma^2+12 \Gamma+5)+\cdots  \qquad\mbox{as}\quad z\to 
+ \infty  , \nonumber\\
v &=& -\frac{\nu (1+ 2 \Gamma)}{2 z}-\frac{1}{8}\frac{\nu^3}{z^4}(4\Gamma^2-1)(2\Gamma+3)+\cdots  \qquad\mbox{as}\quad z\to - \infty  ,\nonumber\\
{\bar v} &=& - z^{1/2}+\frac{1}{4}\frac{\nu (1 - 2\Gamma)}{z}+\frac{1}{32}\frac{\nu^2}{z^{5/2}}(12\Gamma^2-12 \Gamma+5)+\cdots  \qquad\mbox{as}\quad z\to 
+ \infty  , \nonumber\\
{\bar v} &=& -\frac{\nu (1- 2\Gamma)}{2 z}+\frac{1}{8}\frac{\nu^3}{z^4}(4\Gamma^2-1)(2\Gamma-3)+\cdots  \qquad\mbox{as}\quad z\to - \infty .
  \label{eq:formsofv}
\end{eqnarray}
The interpretation of these terms is that they all have at least one boundary ending on the FZZT brane. The Euler number is therefore modified to $\chi = 2 -2h-b-f$, where $f$ is the number of FZZT boundaries. So the first term in the positive $z$ $v(z)$ expansion is the FZZT disc, and the next two are cylinders, the first with both boundaries on the FZZT brane, the second with one boundary on the FZZT brane and one on the ZZ branes. The $\bar{v}$ expansion can be obtained from the $v$ expansion by multiplying the coefficient of each surface by $(-1)^f$.
\\

Another property of the string equation \reef{eq:DJM} is that solutions, $u_{\Gamma}(z)$, with a given value of $\Gamma$ can be related to the solutions $u_{\Gamma \pm 1}(z)$ by the KdV \emph{B\"acklund transformation}:
\begin{eqnarray} \label{eq:Backlund}
w_{\Gamma}^\prime + w_{\Gamma \pm 1}^\prime = \frac12 (w_{\Gamma \pm 1}-w_{\Gamma})^2 \, , \qquad u_\Gamma = w^\prime_{\Gamma} \quad u_{\Gamma \pm 1} = w^\prime_{\Gamma \pm 1}   .
\end{eqnarray}
The physics of this equation when applied to string theory has been studied in great detail in \cite{Dalley:1992br, Seiberg:2004ei, Carlisle:2005mk,Carlisle:2005wa}.
\\


\section{Supercritical String Theory?} \label{eq:Super}

In this section we will attempt to extend the results of the last section to negative values of $m$. If we take \reef{eq:Supercharge2} at face value then we expect to find that we have left the subcritical $\hat{c}<1$ domain and have instead broken through to the the $\hat{c} \geq 9$ domain (Fig. \ref{fig:cplot}). To do this would obviously require some sort of formal analytic continuation. Such a continuation has been speculated before, such as in \cite{Gross:1989vs} for instance, but never successfully carried out in the string theory context. We would like to find new string equations for the supposed $(2, -4|m|)$ models. Looking at the form of the recursion relations \reef{eq:StringRecursion} this does not seem feasible.

However, let us reformulate the KdV hierarchy in a way that will allow us to more easily define any analytic continuation. One can write the $m$--th member of the KdV hierarchy in the following form \cite{Dickey}:
\begin{eqnarray} \label{eq:KdvOther}
\alpha_m \frac{\partial u(x,t_m)}{\partial t_m} = R^\prime_{m+1}[u] \equiv K^m \cdot u' \, , \qquad K \equiv \partial^2 - 4 u - 2 u' \int_x ,
\end{eqnarray}
where $\int_x$ is an operator that integrates with respect to $x$. As was demonstrated in \cite{Carlisle:2005mk}, one way to get from the KdV equation to the string equation is to search for scaling solutions of the former equation. To do this we write $u(x,t_m) = t_m^{2 \beta_m} u(z)$, with $z = x t_m^{\beta_m}$, and $\beta_m$ a constant that will be determined below. Substituting into \reef{eq:KdvOther} we find:
\begin{eqnarray} 
t_m^{2 \beta_m -1} \alpha_m \beta_m (2 u + \tilde{z} u') = t_m^{(2m+3) \beta_m} K^m \cdot u'  , \qquad K \equiv \ud^2 - 4 u - 2 u' \ud^{-1}  ,
\end{eqnarray}
where $\ud^{-1}$ refers to integration with respect to $z$ (with an appropriate $\nu^{-1}$ factor). We see that to eliminate $t_m$ from this equation we need to choose $\beta_m = -1/(2m+1)$. We obtain:
\begin{eqnarray} \label{eq:FA1} 
\alpha_m \beta_m (2 u + \tilde{z} u') = \tilde{\alpha}_m \nu K \cdot 1 =  K^m \cdot u'  ,
\end{eqnarray}
for some constant $\tilde{\alpha}_m$. Notice that we are always entitled to leave in an integration constant when we act with the operator $K$. It is clear that this will mix $K^m$ with lower powers of $K$ in any combination we wish. This allows us to modify \reef{eq:FA1} to:
\begin{eqnarray} \label{eq:FA2}
\sum_{n=1}^m \tilde{t}_n K^n \cdot u' - K \cdot \nu = 0 \quad &\Rightarrow& \quad K \cdot \left( \sum_{n=0}^{m-1} \tilde{t}_{n+1} K^n \cdot u' - \nu \right) = K \cdot \mathcal{R}^\prime = 0 , \\ \label{eq:Rdef} \mathcal{R}^\prime &\equiv& \sum_{n=0}^{m-1} \tilde{t}_{n+1} K^n \cdot u' - \nu  ,
\end{eqnarray}
where the $\tilde{t}_n$ are arbitrary constants. Comparison of \reef{eq:Rdef} to \reef{eq:Bugg} confirms that they are the same up to rescaling of the $\tilde{t}_n$. To proceed we now multiply \reef{eq:FA2} by $\mathcal{R}$ and integrate. Upon doing this we once again obtain the string equation:
\begin{eqnarray} \label{DJMagain}
u{\cal R}^2-\frac{1}{2}{\cal R}{\cal R}^{''}+\frac{1}{4}({\cal R}^{'})^2 =\nu^2\Gamma^2 .
\end{eqnarray}
\\

\subsection{The Negative KdV Hierarchy and its String Equations}

As promised, we will now analytically continue the KdV equation to negative values of $m$. To do this is extremely simple: we just act on \reef{eq:KdvOther} by $K^{-m}$ on both sides. This defines the \emph{negative KdV hierarchy}, which is well known in the mathematical literature \cite{Hone:1999}. The first member of the series is related by a change of variables to the \emph{Camassa-Holm} equation \cite{Hone:1999}. Using the above scaling we see that we should be able to derive the negative $m$ analogues of the string equation. Using \reef{eq:FA1} we find that we can write:
\begin{eqnarray} \label{eq:NegKdv}
K^{|m|+1} \cdot \nu = \kappa_{|m|} u'  ,
\end{eqnarray}
for some constant $\kappa_{|m|}$. In general this will be an integro-differential equation rather than a mere differential equation, so obtaining solutions may be difficult. However in the case of the $m=-1$ model, a substitution will render it in the form of a pure differential equation. More generally we can again include integration constants in $K$. When we do this we find that the $m$--th equation gains contributions from the positive KdV hierarchy:
\begin{eqnarray} \label{eq:NegKdvT}
K^{|m|+1} \cdot \nu + \sum_{n=0}^{|m|-1} \hat{t}_n K^{n} \cdot u' = \kappa_{|m|} u'  .
\end{eqnarray}
In the positive $m$ models the constants $\tilde{t}_n$ correspond to coefficients of operators in the worldsheet conformal field theory. It is plausible that the $\hat{t}_n$ play a similar role in the negative $m$ models. It is therefore sensible to set the $\hat{t}_n$ to zero for the time being and to try to understand the unperturbed models first. We will briefly discuss what happens when they are non-zero later on. However, it is interesting to note that we can always choose a value of $\hat{t}_0$ that allows us to eliminate $\kappa_{|m|}$. 
\\

Before attempting any analysis of these new string equations, let us recall that they could potentially describe the supercritical $(2, -4|m|)$ models, if such models exist. If the equations yield any physically sensible results at all then this would be highly non-trivial. Should this be the case then there would remain the question of whether we have indeed broken through the $\hat{c}=1$ barrier and reached $\hat{c} > 9$? Or whether we have `bounced off' and are now studying some other $\hat{c} < 1$ model? We will leave these questions until Section \ref{sec:Discuss}. Regardless of whether we are dealing with supercritical models or not, we will still refer to them as such in what follows to distinguish them from the earlier subcritical models.
\\

We will later write the negative KdV string equations in their natural form \reef{DJMagain}, but for now let us analyse \reef{eq:NegKdv} in its present form. We will see that this will yield valuable information that would otherwise have been obscured. The first member of the hierarchy is $m=-1$. By writing $u(z) = w^\prime (z)$ we obtain the following differential equation:
\begin{eqnarray} \label{eq:Neg1String}
4 w^{(3)} + \tilde{z} w^{(4)} - 8 (w')^2 - 6 \tilde{z} w' w'' - 2 w w'' = \kappa_1 w''  . 
\end{eqnarray}
If this equation describes the $(2,-4)$ superconformal model as our analytic continuation suggests, then it would correspond to $\hat{c}=10$. So this would be eleven dimensional Type 0A superstring theory, with ten `ordinary' dimensions plus the timelike Liouville mode. The first observation of note is that the constant $\kappa_1$ can be eliminated by adding a constant term to $w(z)$. This term will be an analytic contribution to the free energy and as such is non--universal so can be ignored. So we can set $\kappa_1=0$ without loss of generality. 
\\

Recalling the subcritical models, we should naturally expect the leading order behaviour of $u(z)$ to be $z^{1/m}$ for the $(2, 4m)$ model. So for $m=-1$ we expect $w(z)$ to lead with a $\textrm{ln} \,z$ contribution.  This has a problematic interpretation and leads to terms of the form $z^a \textrm{ln} \, z$ in the free energy. It turns out that the $\textrm{ln} \,z$ ansatz is not a solution of the equation anyway, so we are saved from having to make such an interpretation. So on first glance it looks as if this cannot possibly be the $(2, -4)$ model. This is not necessarily the case however, and to show this we need to understand the origin of the the expected $1/m$ power. This comes from KPZ scaling. In the case that $z$ is the bulk cosmological constant one expects the first term in the free energy to lead like $z^{-\gamma_{str}}$, where $\gamma_{str}$ is given by \cite{Ginsparg:1993is, Distler:1989nt}:
\begin{eqnarray}
\gamma_{str} &=& 2 + \frac{1}{12} \left( c-25 - \sqrt{(25-c)(1-c)} \right)  ,  \\
\gamma_{str} &=& 2 + \frac{1}{4} \left( \hat{c}-9 - \sqrt{(9-\hat{c})(1-\hat{c})} \right)  ,
\end{eqnarray}
where we use the $c$ formula in the bosonic theory; and the $\hat{c}$ formula in the Type 0A theory. In the bosonic theory one sees that the unitary $(m,m+1)$ models have an expected value of $\gamma_{str} = -1/m$. So the $m=3$ critical point is naively expected to be the $(3,4)$ Ising model. However, studies have revealed \cite{Staudacher:1989fy} that this model actually corresponds to the $(2, 5)$ Yang-Lee edge singularity, which has a different predicted value of $\gamma_{str}$. The resolution to this problem is to speculate that the coupling of the conformal field theory to Liouville gravity has introduced a negative dimension operator to which $z$ couples. This means that $z$ is no longer the bulk cosmological constant in these non-unitary models\footnote{We stated this in the above, but here is the explanation.}. It was this argument which originally led to the conclusion that the $m$--th multicritical point corresponded to the $(2, 2m-1)$ model, and not the $(m, m+1)$ model as was originally believed. So we should not be overly concerned if we do not obtain a $z^{1/m}$ scaling law for our new models: the continuum scaling results did not even hold in the subcritical models in the first place.
\\

We proceed by looking for a leading order solution of the form $w(z) \sim B z^b$, for some power $b$ and coefficient $B$. Substituting into \reef{eq:Neg1String} we find that there are two types of contribution: terms proportional to $z^{b-3}$, and those proportional to $z^{2b-2}$. The $B z^b$ ansatz will be appropriate for most equations that are similar in form to \reef{eq:Neg1String}. Accordingly we will explain how to proceed in this general case, using the specific example of \reef{eq:Neg1String} to illustrate matters. In most cases there will be two ways to fix a $B z^b$ term in the perturbative expansion; and we will call these \emph{power determined} and \emph{coefficient determined}. 

The first involves simply solving for when two or more terms become equally dominant. In the specific case of \reef{eq:Neg1String} this means that we solve the equation $b-3=2b-2$. In the general case one can plot all the contributing terms as lines on a graph, with $b$ on the horizontal axis, and the powers of $z$ on the vertical axis. One then looks for dominant crossing points on the graph by starting on the top line at $b=-\infty$ and tracing down until one reaches an intersection of two or more lines. One can continue in this way, always staying on the highest line until $b=+\infty$ is reached. Once all the critical values of $b$ have been power determined in such a way it is possible to determine the corresponding values of $B$. We call this power determined because $b$ depends only on the general form of the differential equation, not the numerical factors weighting the different terms themselves. This method was employed exclusively in calculating the expansions in Section \ref{sec:Sub}.

If two or more terms have the same power in an equation then one may find $b$ another way: the terms form a subgroup that vanishes if $b$ takes on special values. This will be true for all values of $B$, and hence corresponds to the introduction of an arbitrary constant into the solution. We call this coefficient determined, because $b$ will depend on the numerical coefficients in the differential equation, and not just the general form of that equation.
\\

In the case of \reef{eq:Neg1String} we find that $b=-1$ is the only power determined value. For $b<-1$ the $z^{b-3}$ terms will dominate; and for $b>-1$ the $z^{2b-2}$ terms will dominate. The $z^{b-3}$ subgroup will vanish if $b$ satisfies the following equation:
\begin{eqnarray}
b(b-1)(b-2)[4 + (b-3)] = 0  .
\end{eqnarray}
So it will vanish for $b=-1,0,1,2$. However, all these values apart from the first lead to subleading terms ($b>-1$) and so can be discounted. The $z^{2b-2}$ subgroup vanishes if:
\begin{eqnarray}
8 b^2 + 6 b^2 (b-1) - 2 b (b-1) = 0  .  
\end{eqnarray}
So it will vanish for $b=-1, 0, 1/3$. The $b=-1$ term is common to both the $z^{b-3}$ and $z^{2b-2}$ subgroups and hence will override the earlier power determined value to give us an arbitrary coefficient of a $z^{-1}$ term. We will return to this later. The $b=0$ solution is just the trivial constant that we eliminated before. So the only new value is $b=1/3$. We write $w(z) = A^2 z^{1/3}/(3\nu) + w(z)$ for some free parameter $A$ (why we have written it in this way will become clear shortly). We then solve the resulting equation in the same way for the new $w(z)$. There is an arbitrary constant at the next order too. We will call it $\Gamma$. The first few terms of the solution are:
\begin{eqnarray} \label{eq:Soln1}
w = \frac{A^2 z^{1/3}}{3 \nu} + \frac{2 A \Gamma}{3 z^{1/3}} - \frac{\nu (12 \Gamma^2-5)}{36 z} + \frac{2 \nu^2 \Gamma (\Gamma^2-1)}{9 A z^{5/3}} - \frac{\nu^3 \Gamma^2 (\Gamma^2-1)}{9 A^2 z^{7/3}} + \cdots  .
\end{eqnarray}
Integrating up once (and dividing by $\nu$) to get the free energy, we find:
\begin{eqnarray}
F = \frac{A^2 z^{4/3}}{4 \nu^2} + \frac{A \Gamma z^{2/3}}{\nu} - \frac{(12 \Gamma^2-5)}{36} \textrm{ln} \, z - \frac{\nu \Gamma (\Gamma^2-1)}{3 A z^{2/3}} + \frac{\nu^2 \Gamma^2 (\Gamma^2-1)}{12 A^2 z^{4/3}} + \cdots  .
\end{eqnarray}
We see that once again we can define a dimensionless parameter $g_s=\nu/(A z^{2/3})$, though this time it contains our constant $A$. We can therefore write the free energy as:
\begin{eqnarray} \label{eq:MoreFree}
F = \frac{1}{4} g_s^{-2} + \Gamma g_s^{-1} - \frac{(12 \Gamma^2-5)}{36} g_s^0 \ \textrm{ln} \, z - \frac{\Gamma (\Gamma^2-1)}{3} g_s + \frac{\Gamma^2 (\Gamma^2-1)}{12} g_s^2 + \cdots  .
\end{eqnarray}
The interpretation of these terms can again be made in terms of worldsheets with some number of handles and boundaries. $\Gamma$ would then represent the number of a species of background branes. Is it the same $\Gamma$ that we found in the positive $m$ hierarchy? Let us evaluate the B\"acklund transformation \reef{eq:Backlund}. We should ask whether we expect the B\"acklund transformation to hold at all in these new models? The answer is yes, for the following reason. The Lax equations for the first member of the negative KdV hierarchy are given by \cite{Hone:1999}:
\begin{eqnarray}
\left[ \partial^2 - u(x,t) \right] \psi = \lambda \psi \, , \qquad w \frac{\partial \psi}{\partial x} - \frac{1}{2} \frac{\partial w}{\partial x} \psi = 2 \lambda \frac{\partial \psi}{\partial t} \, , \qquad \quad u = \frac{\partial w}{\partial x}  . 
\end{eqnarray}
The first of these equations is identical to \reef{eq:WF1}, implying that we can once again derive the Miura map \reef{eq:Miura}. The B\"acklund transformation \reef{eq:Backlund} is derived from the Miura map by noticing that if the mKdV hierarchy has a solution $v(z)$ then it also has a solution $-v(z)$. The mKdV hierarchy can be formulated in a similar manner to \reef{eq:KdvOther}. If we do this then it becomes clear that the property in question should still be true for the negative mKdV hierarchy. So we expect that the B\"acklund transformation should still be relevant in our new models. Plugging the solution \reef{eq:Soln1} into \reef{eq:Backlund}, we find that the transformed expansion is indeed of the same form as \reef{eq:Soln1}. What is more, $\Gamma$ has indeed changed by an integer. This is evidence that the model may be describing a valid physical theory. 
\\

Notice also that the solution \reef{eq:Soln1} appears to have special properties if $\Gamma=0,\pm1$. We calculated the expansion up to tenth order and found that, after the torus term, all the higher order contributions are proportional to $\Gamma(\Gamma^2-1)$ multiplied by some other polynomial in $\Gamma$. This means that they vanish at these special values of $\Gamma$. As alluded to above, there is also a second large--$z$ solution to \reef{eq:Neg1String} that leads with $w(z) \sim Cz^{-1}$, where $C$ is a constant. It turns out that the second term in the expansion is also coefficient determined. What is more, the condition for the relevant subgroup to vanish now contains $C$ itself. So the powers appearing in the series actually depend on our choice of the leading term. This is something that has no analogue in the subcritical models. We find that if the next leading order term is of the form $z^b$ then $b$ must satisfy:
\begin{eqnarray}
b = 1 \pm \sqrt{1 - 4C}  .
\end{eqnarray}
Looking back at \reef{eq:UExpandNeg} we see that the first term in that expansion also contributes a $z^{-1}$ contribution to $w(z)$. In fact, for the subcritical models this first term is shared by all the string equations in the hierarchy. The coefficient is the same too. So let us speculate that this first term is shared by all the members of the negative hierarchy as well. Writing $C=-(4\Gamma^2-1)/4$ we find that $b$ simplifies to give $b = 1 \pm 2 \Gamma$. Assuming for now that $\Gamma$ is positive and greater than unity, we take the negative sign choice in this equation for $b$. We then find that the following perturbative solution holds at least up to tenth order:
\begin{eqnarray} \label{eq:Soln2}
w &=& -\frac{\nu(4 \Gamma^2-1)}{4 z} + \sum_{n=1}^\infty \frac{(-1)^{n-1} \nu^{n+1} B^{n} }{4^{n-1} (\Gamma-1)^{n-1}} z^{2n-1-2n\Gamma} , \\
F &=& -\frac{4 \Gamma^2-1}{4} \textrm{ln} \, z  +  \sum_{n=1}^\infty \frac{(-1)^{n} \nu^n B^{n}}{2^{2n-1} n (\Gamma-1)^n} z^{-2n(\Gamma-1)}  .
\end{eqnarray}
Here $B$ is another constant. Assuming that the dimensionless string coupling is of the same form, $g_s=\nu/(A z^{2/3})$, as in \reef{eq:MoreFree} \footnote{Though there is no conclusive reason why this needs to be the case.}, we can rewrite the above free energy in the following form:
\begin{eqnarray} \label{eq:FreeNegGam}
F = -\frac{4 \Gamma^2-1}{4} \textrm{ln} \, z  +  2 \sum_{n=1}^\infty \frac{\tilde{B}^n}{n} g_s^{3 n(\Gamma-1)}  ,
\end{eqnarray}
where we have absorbed all extra $\Gamma$ and $\nu$ dependence into the new constant $\tilde{B}$. The physical interpretation of this series is indeed curious. The surfaces contributing to the free energy depend on the value of $\Gamma$. What is more, it seems that we must have $\Gamma>1$ to have $\tilde{B}$ non-zero\footnote{Recall that we chose $b=1-2\Gamma$. If instead we choose $b=1+2\Gamma$ then we can have $\Gamma<-1$.}. It is also intriguing that we seem to have obtained a general form for the series to all orders in perturbation theory. Many of these mysteries are solved by the B\"acklund transformation, under which it turns out that one must set $\tilde{B}=0$ to make $\Gamma$ change by an integer. If we take the B\"acklund transformation as in some way sacrosanct to the physics of the theory, then we conclude that the physical theory will always have $\tilde{B}=0$, thus eliminating the need to explain this apparently troublesome physics. It is nonetheless an interesting solution and perhaps there will turn out to be a role for it. 
\\

Note also, that unlike in the $m=1$ model discussed earlier, both of the possible large--$z$ expansions, \reef{eq:Soln1} and \reef{eq:Soln2} (with $\tilde{B}=0$), can be real at both large negative and large positive $z$. So it is perhaps possible that we can have \reef{eq:Soln1} or \reef{eq:Soln2} being valid in both weak coupling regimes. We will see a useful interpretation of this a little later. Unfortunately we have not yet been able to solve the string equation of \reef{eq:Neg1String} numerically. If this could be done then it would yield vital information about which boundary conditions we should choose. For now let us consider the Miura map \reef{eq:Miura}, and calculate what we would naively conclude are the FZZT free energies corresponding to \reef{eq:Soln1} and \reef{eq:Soln2}. We could leave $\tilde{B} \neq 0$ in the latter case, but this is not very insightful. For \reef{eq:Soln1} we again find two solutions $v(z)$ and $\bar{v}(z)$: 
\begin{eqnarray}
v &=& \frac{A}{3 z^{1/3}} - \frac{\nu(2 \Gamma + 1)}{6 z} + \frac{\nu^2 \Gamma(\Gamma+1)}{3 A z^{5/3}} - \frac{\nu^3 \Gamma (2\Gamma^2 + 3\Gamma+1)}{9 A^2 z^{7/3}} + \cdots  , \nonumber\\
{\bar v} &=& -\frac{A}{3 z^{1/3}} + \frac{\nu(2 \Gamma - 1)}{6 z} - \frac{\nu^2 \Gamma(\Gamma-1)}{3 A z^{5/3}} + \frac{\nu^3 \Gamma (2\Gamma^2 - 3\Gamma+1)}{9 A^2 z^{7/3}} + \cdots  . 
\end{eqnarray}
Turning to the solutions of the Miura map associated with \reef{eq:Soln2} we find:
\begin{eqnarray}
v = -\frac{\nu (1+ 2 \Gamma)}{2 z}  , \qquad
{\bar v} = -\frac{\nu (1- 2\Gamma)}{2 z}  ,
\end{eqnarray}
which are exact solutions. Once again we can interpret these results in terms of open strings having $f$ boundaries ending on a new type of brane. Notice that these $v(z)$ and $\bar{v}(z)$ expansions respect the $(-1)^f$ symmetry of the subcritical models. If these models are indeed related to supercritical string theories then it is questionable whether we should expect to find the physics given in terms of ZZ and FZZT branes like in the subcritical models. Here we merely point out what the mathematics is telling us. We will postpone discussion of the physics until Section \ref{sec:Discuss}. 
\\

\subsection{Higher String Equations}

Let us study the next string equation in the hierarchy, $K^3 \cdot \nu = 0$, which corresponds to the supposed $(2,-8)$ model. It is an integro-differential equation with the naive value of central charge $\hat{c}=27/2$:
\begin{eqnarray} \label{eq:HighString}
&& 6 w^{(5)} + \tilde{z} w^{(6)} - 36 (w'')^2 - 48 w'w^{(3)} - 20 \tilde{z} w'' w^{(3)} - 10 \tilde{z} w' w^{(4)} - 2 w w^{(4)} + 32 (w')^3 \nonumber \\ &+& 24 \tilde{z} (w')^2 w'' + 8 w w' w'' + 4 w'' \int \ud z \, \left( 4 (w')^2 + 3 \tilde{z} w' w'' + w w'' \right) = \kappa_{2} w''   ,
\end{eqnarray}
where as usual the integration comes with a factor of $\nu^{-1}$. With $\kappa_{2}$ non-zero it appears that there are no perturbative solutions to this equation at all. So it seems as if we are forced to set the $\kappa_{|m|}$ to zero to proceed. When we do this we realise that the negative KdV string equations have the form $K^{|m|+1} \cdot \nu = 0$. So a solution of the $|m|$--th model will also be a solution of the $(|m|+1)$--th. There is one new solution to \reef{eq:HighString} though:
\begin{eqnarray}
w = \frac{A^2 z^{3/5}}{15 \nu} + \frac{2 A \Gamma}{5 z^{1/5}} - \frac{\nu(12 \Gamma^2 - 7)}{60 z} + \frac{\nu^2 \Gamma (12 \Gamma^2-18)}{45 A z^{9/5}} - \frac{\nu^3 (18 \Gamma^4 - 42 \Gamma^2 +7)}{45 A^2 z^{13/5}} + \cdots  .
\end{eqnarray}
To this expansion we can assign the dimensionless string coupling $g_s = \nu/(A z^{4/5})$. Once again, the parameter $\Gamma$ behaves as expected under the B\"acklund transformation \reef{eq:Backlund}. In general it seems that the $|m|$--th string equation of the negative KdV hierarchy has a solution $u(z)$ with $z^{-2/(2|m|+1)}$ leading order behaviour. This is in addition to it having all the solutions of the lower equations. Notice that we can also define the $m=0$ model in \reef{eq:NegKdv}. It is to this model that the $u(z)= (4 \Gamma^2-1)/(4z^2)$ solution most naturally belongs. Unfortunately, we see from \reef{eq:Supercharge2} that the naive value of the central charge for this solution is infinite. This will be discussed further in Section \ref{sec:Discuss}.
\\

Since all the perturbative solutions are valid at both negative and positive large--$z$, we seem to have a lot of choices of boundary conditions with no obvious way to choose between them. However, so far we have been dealing with the differentiated analogue of the string equation \reef{eq:DJM}. That is, we have been studying $K \cdot \mathcal{R}^\prime = 0$. This we have done for reasons that will become clear below. A formulation of the string equations analogous to \reef{eq:DJM} is possible however, and we will derive it now. This will clarify a few issues and lead to some important restrictions upon the solutions we have already obtained.
\\

\subsection{The Full String Equation}

Recall our basic string equation \reef{eq:NegKdv}:
\begin{eqnarray}
K^{|m|+1} \cdot \nu = \kappa_{|m|} u'  .
\end{eqnarray}
We can integrate up to get the analogue of \reef{eq:DJM} if we can write it in the form $K \cdot \mathcal{R}^\prime = 0$. We see that if $\kappa_{|m|}=0$ then we can achieve this in closed--form:
\begin{eqnarray}
K \cdot \left( K^{|m|} \cdot \nu \right) = 0 \quad \Rightarrow \quad \mathcal{R}^\prime = K^{|m|} \cdot \nu  .
\end{eqnarray}
Earlier it was explained how leaving $\kappa_{|m|}$ non-zero caused problems when searching for perturbative solutions. We also saw how it could always be absorbed into $\hat{t}_0$ in \reef{eq:NegKdvT}. So although it may be possible to leave $\kappa_{|m|}$ non-zero and still compute an appropriate $\mathcal{R}$, it seems reasonably natural to set it to zero. This done, $\mathcal{R}$ satisfies the following equation:
\begin{equation} \label{eq:DJM2}
u{\cal R}^2-\frac{1}{2}{\cal R}{\cal R}^{''}+\frac{1}{4}({\cal R}^{'})^2 =\nu^2 \Lambda^2  ,
\end{equation}
which is the usual string equation of the positive $m$ models as we had hoped. $\Lambda$ is a constant that we suspect will be related to $\Gamma$. For the $m=-1$ model we find that $\mathcal{R}$ is given by:
\begin{eqnarray}
\mathcal{R} = -2\nu \left[ w(z) + \tilde{z} w'(z) \right]  ,
\end{eqnarray}
which yields the full string equation:
\begin{eqnarray} \label{eq:NewStringDJM}
&& w^2 w' + 2 \tilde{z} w (w')^2 + \tilde{z}^2 (w')^3 - \frac{3}{2} w w'' - \frac{1}{2} \tilde{z} w w^{(3)} \nonumber \\ &-& \frac{1}{2} \tilde{z} w' w'' - \frac{1}{2} \tilde{z}^2 w' w^{(3)} + (w')^2 + \frac{1}{4} \tilde{z}^2 (w'')^2 - \frac{\Lambda^2}{4} = 0  .
\end{eqnarray}
We can substitute our original perturbative solutions into this equation to see if they are still valid. Starting with \reef{eq:Soln1} we uncover the following, perhaps surprising, result: the expansion actually satisfies \reef{eq:NewStringDJM} for \emph{all} values of $\Gamma$. So it seems that $\Gamma$ and $\Lambda$ are independent parameters. Instead we find that it is actually $A$ that is related to $\Lambda$:
\begin{eqnarray}
\nu \Lambda = \pm \left( \frac{2 A}{3} \right)^3  .
\end{eqnarray}
Analogous relations can be found in the other members of the hierarchy\footnote{The exception being the $m=0$ model, which has $\mathcal{R}=z$. In this model $\Gamma$ and $\Lambda$ \emph{are} the same parameter.}. So it has transpired that our constants $A$ and $\Lambda$ are not independent in the context of \reef{eq:DJM2}. We also see that $\Gamma$ does not appear explicitly in either version of the string equation, but instead it can always be found via the method of coefficient determination and the use of the B\"acklund transformation \reef{eq:Backlund}. This is why we studied the differentiated equation first: it is much easier to identify the correct physical $\Gamma$.

However, in this framework the expansion given by \reef{eq:Soln1} is clearly singular if $\Lambda=0$ (because $A=0$). Moreover, the expansion given by \reef{eq:Soln2} is not a solution of \reef{eq:NewStringDJM} \emph{unless} $\Lambda=0$. The same appears to be true of the $|m|$--th model: for $\Lambda\neq0$ the theory is forced to be in its `natural' state with $u(z) \sim z^{-2/(2|m|+1)}$ boundary conditions in both weak coupling regimes; but when $\Lambda=0$ it must drop down to a solution of a lower member of the hierarchy. This is because if $\Lambda$ is explicitly zero in \reef{eq:DJM2} then we can choose $\mathcal{R} \equiv \mathcal{R}_{|m|}=0$, implying that the solutions to the differentiated $K \cdot \mathcal{R}_{|m|-1}^\prime=0$ equation of the \emph{next lowest model} will hold. When the theory does drop out of its natural state there seems to be no obvious way of determining which of the lower solutions it will end up at; or what its new values of $\Lambda$ and $\Gamma$ will be when it gets there.
\\

\section{Discussion} \label{sec:Discuss}

In the preceding section we have seen that it is possible to define a family of string equations associated with the negative KdV hierarchy. It is encouraging that the perturbative expansions obtained from these string equations can again be interpreted sensibly in terms of string worldsheets with various numbers of handles and boundaries. What is more, it seems that we can once again identify a parameter $\Gamma$ that changes by an integer under the B\"acklund transformation \reef{eq:Backlund}. It is of course very possible that we have merely uncovered a mathematical structure that is mimicking sensible physics. However, it is interesting that $\Gamma$ still plays the same role as in the positive KdV string equations, yet no longer appears explicitly in the string equation \reef{eq:DJM2}. 

Assuming for now that the models do describe physical theories, there remains the question of what these theories actually are. Our naive expectation has always been that they are the supercritical $(2,-4|m|)$ models, with central charges $\hat{c} = 10, \ 27/2, \ 52/3, \dots$ ($m=-1,-2,-3,\dots$ respectively). We see from \reef{eq:Supercharge2} that for $m \lesssim -2$ the central charge is well--approximated by $\hat{c}=4|m|+5$, and hence the dimensionality increases almost linearly with $|m|$. However, if the hierarchy really is describing supercritical string theories, then it seems strange that we are again seeing the physics described in terms of objects that look very similar to the ZZ and FZZT branes of the subcritical models. In supercritical string theory the Liouville direction is timelike and so the locations of such branes become problematic. Nevertheless, some progress has been made on these issues, such as in \cite{Gutperle:2003xf, Strominger:2003fn, Fredenhagen:2004cj}, where results from the usual spacelike Liouville theory are analytically continued to obtain sensible answers. So if the negative KdV hierarchy is indeed describing some aspect of supercritical string theories, then perhaps we should not be too surprised that we are again seeing what look like ZZ and FZZT branes.

One would also expect, perhaps naively, that supercritical string theories would be far richer in nature than their subcritical counterparts. Yet the negative KdV models do not appear to be a great deal more complicated than those of the positive hierarchy. So it is perhaps possible that we are not dealing with the supercritical models that we naively think we are, and are instead dealing with some new subcritical models. We should always be careful however, because we are lacking an obvious target spacetime interpretation. Until such an interpretation can be made it is hard to draw any firm conclusions as to the nature of the physical theory.
\\

Recall that each member of the subcritical $(2,4|m|)$ series of string equations has two large--$z$ solutions. One of these corresponds to a weak coupling regime with D-branes, the other to a regime with fluxes. In contrast, the $m$--th string equation of the negative KdV hierarchy \reef{eq:DJM2} has only one perturbative solution for $\Lambda$ non-zero. This has $z^{-2/(2|m|+1)}$ leading order behaviour and can be real--valued in \emph{both} weak coupling regimes. When $\Lambda$ is set to zero we find that this solution no longer exists and instead we are forced choose one of the solutions associated with the lower members of the hierarchy. An interesting interpretation of this is that the lower solutions could correspond to different phases of the \emph{same} theory. 

In \cite{Hellerman:2004zm, Hellerman:2004qa} it was demonstrated that certain heterotic supercritical string theories are unstable and can suffer tachyon condensation down to lower dimensional string theories. This is certainly analogous to what we may be seeing here: string theories of lower dimensionality existing as phases of higher dimensional theories. Perhaps when we set $\Lambda$ to zero we are somehow forcing the theory to condense down to a lower dimensional theory? If so then it would be interesting to understand the physical interpretation of $\Lambda$, and to uncover a dynamical mechanism for how this tachyon condensation works. However, we should again be careful, because the endpoint of this condensation process could well be the $m=0$ model corresponding to the $u(z)= (4 \Gamma^2-1)/(4z^2)$ solution. Recall that this is naively the $(2,0)$ model and hence has infinite central charge. The string theory interpretation of this is unclear, but it would seem to correspond to an infinite dimensional theory. From the point of view of two-dimensional worldsheet gravity it corresponds to taking a semiclassical limit in which the partition function is dominated by stationary points of the action \cite{Polchinski:1989fn, Cooper:1991vg}. Perhaps this is a sign of a fundamental sickness in the models. Or perhaps if we better understood the dynamical mechanism by which the condensation process worked in these theories then we would be able to rule out the $m=-1$ to $m=0$ transition on physical grounds. In \cite{Hellerman:2004zm, Hellerman:2004qa} it was speculated that the endpoint of the condensation process in the heterotic string would be the critical theory. This certainly does not seem to be the case in our models, and so whether this really is a manifestation of tachyon condensation remains to be seen. 
\\

Whilst there are still many unanswered questions, it is very encouraging that the results obtained from the negative KdV string equations are even remotely sensible. It remains to be seen whether they are just a curious mathematical coincidence or something more physical. If they do describe some aspect of supercritical string theory then it would be very exciting indeed. Certainly the way the theory can apparently change its dimension, possibly via some sort of tachyon condensation, is intriguing. In future work it would be insightful to uncover the meaning of the $\hat{t}_n$ coefficients in \reef{eq:NegKdvT}. Are they once more the coefficients of certain primary operators in the theory? We performed some perturbative calculations which showed that the $\hat{t}_n$ do appear in a form that would be consistent with this conjecture. Finally, it would be very interesting to see if we could analytically continue still further to study models with fractional values of $m$. Intuitively though, one feels that this would not be possible.

\section*{Acknowledgements}
This work is supported by an EPSRC studentship at the University of
Durham. JEC thanks Clifford Johnson, Mukund Rangamani, Jos\'e S\'anchez-Loureda, Douglas Smith, Tom Underwood and Andrew Wade for useful discussions. He also thanks the Department of Physics and Astronomy at the University of Southern California for hospitality during the initial stages of this project.

\newpage

\bibliographystyle{utcaps} \bibliography{Supercrit22nd}

\end{document}